\documentclass[aps,prl,showpacs,reprint,superscriptaddress,]{revtex4-1}

\usepackage{graphicx}
\usepackage{dcolumn}
\usepackage{bm}
\usepackage{amsmath, amssymb, mathtools, isomath, nccmath}
\usepackage{siunitx}
\usepackage{xspace}
\usepackage{xcolor}
\usepackage{siunitx}
\usepackage[british]{babel}
\usepackage{sidecap}

\usepackage[colorlinks,
            linkcolor=red,     
            anchorcolor=blue,  
            citecolor=blue,      
            ]{hyperref}

\newcommand{\ket}[1]{\ensuremath{\lvert #1 \rangle}\xspace}%
\newcommand{\bra}[1]{\ensuremath{\langle #1 \rvert}\xspace}%
\usepackage{xcolor}

\begin{document}
 
  \title{
Designing exceptional-point-based graphs yielding topologically guaranteed quantum search}

\author{Quancheng Liu}
  \affiliation{Department of Physics, Institute of Nanotechnology and Advanced Materials, Bar-Ilan University, Ramat-Gan 52900,
  Israel}
\author{David A. Kessler}
  \affiliation{Department of Physics, Bar-Ilan University, Ramat-Gan 52900,
  Israel}
\author{Eli Barkai}
  \affiliation{Department of Physics, Institute of Nanotechnology and Advanced Materials, Bar-Ilan University, Ramat-Gan 52900,
  Israel}

  \date{\today}

  \begin{abstract}

Quantum walks underlie an important class of quantum computing algorithms, and represent promising approaches in various simulations and practical applications. Here we design stroboscopically monitored quantum walks and their subsequent graphs that can naturally boost target searches. We show how to construct walks with the property that all the eigenvalues of the non-Hermitian survival operator, describing the mixed effects of unitary dynamics and the back-action of measurement, coalesce to zero, corresponding to an exceptional point whose degree is the size of the system. Generally, the resulting search is guaranteed to succeed in a bounded time for any initial condition, which is faster than classical random walks or quantum walks on typical graphs. We then show how this efficient quantum search is related to a quantized topological winding number and further discuss the connection of the problem to an effective massless Dirac particle.

  \end{abstract}
\maketitle

Quantum walks \cite{Aharonov1687,Mlken2011}, the quantum analog of the well-known classical random walks, have attracted increasing attention due to its importance both in fundamental physics and applications for quantum information processing \cite{Kempe2003}. Taking advantage of coherent superposition and interference, the quantum walk in many respects is superior to its classical counterpart and finds applications in quantum algorithms \cite{Ambainis2003,Qiang2016}, universal quantum computation \cite{Childs2009,Childs2013}, quantum simulation \cite{MULKEN201137,Xu2021}, and bio-chemical processes \cite{Cao2020,dudhe2021testing}. One main challenge of the quantum walk is to maximize the detection probability on a predetermined target state $\ket{\psi_{\rm d}}$ given some initial state $\ket{\psi_0}$ \cite{Childs2004,Tang2018}. With unitary evolution, nearly perfect quantum search with detection probability approaching unity was found in several graphs for some special initial states at some particular time $t$, including a glued binary tree \cite{Tang2018}, a hypercube, and high-dimensional lattices \cite{Childs2004}, while typical systems fall far from this limit. In a broad sense, the transmission of a known initial state to another state is called quantum state transfer \cite{Cirac1997,Reiserer2015,PhysRevLett.116.100501}. For instance, Kostak {\em et al.} designed permutation operations that propagate the system from one specific node of the graph to another at a predetermined time $t$ \cite{Jex2007}. However, if one does not know what the initial condition is, as is typical in many search problems, we cannot rely on the quantum state transfer or one-shot measurement quantum walks. Therefore, we herein design special graphs and measurement protocols, with the aim to achieve what we call a guaranteed search. Namely, the quantum walker should be successfully detected in a bounded time for any initial condition. We describe how to construct such quantum graphs and corresponding measurement strategies. We further investigate whether these measurements, that destroy the unitary evolution, are either harmful or useful for the search, and in what sense.

 Our work is motivated by the state-of-the-art technology advances in experiments \cite{Daley2022} that allow clever engineering of Hamiltonians with superconducting circuits \cite{ming2021}, waveguide arrays \cite{Mittal2014,Keil2016,Caruso2016,Boada2017,Mittal2019,Chen2021},  trapped ions \cite{Manovitz2020,Monroe2021}, and arrays of neutral atom generated either in an optical cavity~\cite{Periwal2021} or via optical tweezers~\cite{Bluvstein2022}. For instance, using photons carrying information between atomic spins, programmable non-local interactions in an array of atomic ensembles are realized in an optical cavity \cite{Periwal2021}. These advances allow us to consider the option of constructing a device with non-trivial matrix elements of the Hamiltonian and thus design special types of graphs to speed up the quantum search.
 
 We find that the designed quantum graphs, together with the stroboscopic search protocol, have remarkable search capabilities, either with or without control of the initial state. The ability to search an unknown initial state, i.e., a black box initial state, is a significant step forward, in contrast to previous works that considered quantum walks that start from a uniform or specific localized initial state.   Physically, one of the features of efficient quantum search we find here is that it is intimately related to the study of exceptional points. The latter are degenerate eigenvalues of the non-Hermitian operators that are studied, for example, in optics and laser physics \cite{Miri2019,Liertzer2021,Doppler2016,Xu2016}, and topological phases \cite{Lee2016,Leykam2016,Shen2018,Bergholtz2021}, and are fundamentally related to  parity-time symmetry breaking \cite{Guo2009,ElGanainy2018,Xiao2021}. Here, we design graphs and a search protocol with an exceptional point of unusually high degeneracy, namely the size of the entire Hilbert space, which can be made as large as we wish. We highlight the idea that exceptional search is found when all the eigenvalues of the survival operator, defined below, coalesce to zero, creating a large degeneracy. We then explore the topology of the model at the exceptional point and show that efficient quantum search is related to the quantitation of certain topological winding numbers. Towards the end of the paper, we show how the search problem and the corresponding degeneracy of the exceptional points and the topology property are related to an effective massless Dirac particle, though all along we use Schr\"{o}dinger dynamics. We also show how our search strategies are related to quantum state transfer.

\section*{Stroboscopic search protocol and $N$-th order exceptional point}

 To perform efficient quantum walks, we use the strategy of stroboscopic measurements, which as we show later can be made into an efficient tool. In the stroboscopic protocol, the quantum walker starts from an unknown initial state $|\psi_{0}\rangle$ and evolves unitarily according to the graph Hamiltonian $H$.  We projectively measure the system at times $\tau, 2 \tau, \cdots $, at each measurement asking if the system is found at its target, namely at $|\psi_{\rm d}\rangle$. The search target can be a localized node on the graph, however in general this is not a requirement. This yields a string of $n-1$ successive No's followed by a Yes from the $n$-th measurement. Once we record a Yes, the system is at the target state $|\psi_{\rm d}\rangle$ and in that sense, we have a successful quantum search. The time $n \tau$ is the search time of the target state $\ket{\psi_{{\rm d}}}$, which is clearly a random variable whose statistical properties ultimately depend on the initial state of the system $\ket{\psi_{{0}}}$, the unitary evolution between measurements, and the choice of $\tau$. 

\begin{figure}
    \centering
    \includegraphics[width=\columnwidth]{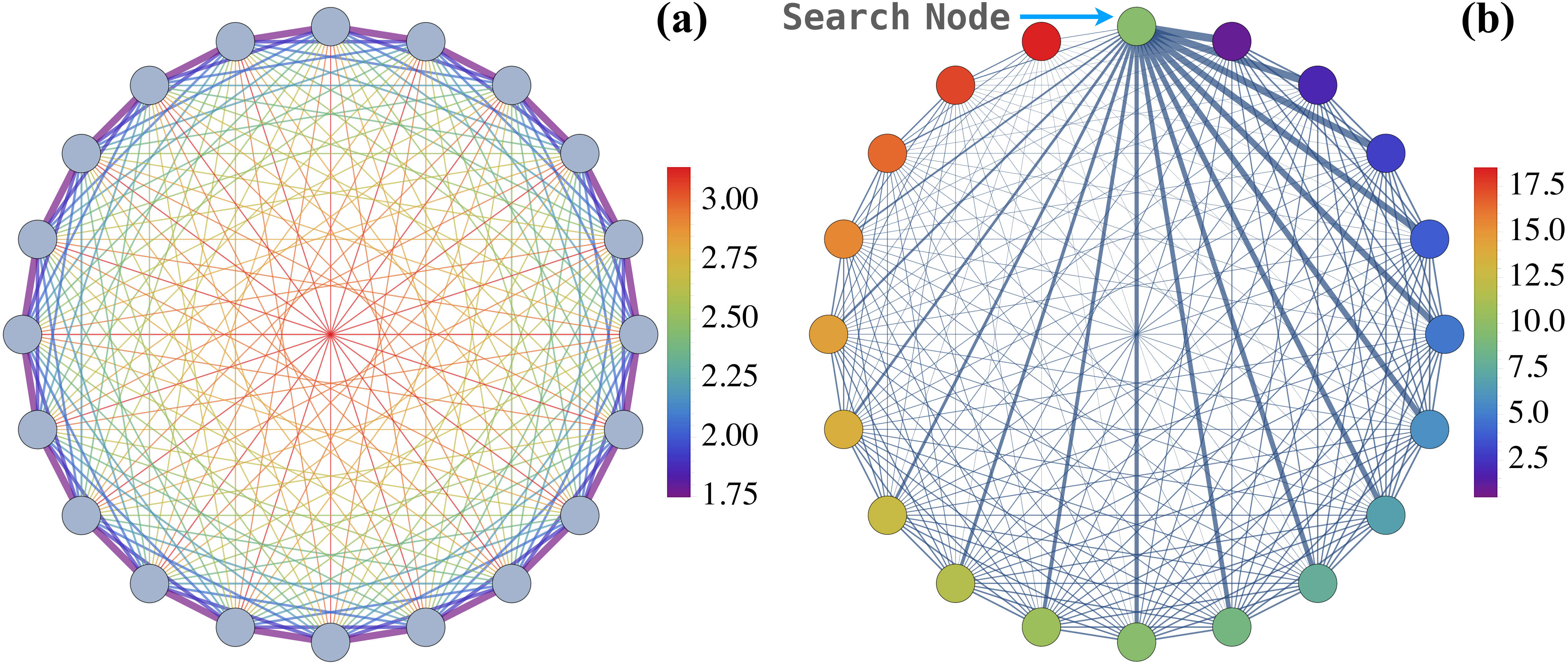}
    \caption{\textbf{Designed quantum graphs.} Schematic presentation of the crawl graph ({\bf a}) and funnel model ({\bf b}). Here $N=20$. The thickness of the connecting line represents the strength of the matrix element connecting two nodes [({\bf a}) and ({\bf b})]. The colors represent the phases of the hopping rates ({\bf a}). In ({\bf b}) we utilize colors to represent the magnitude of the on-site energies, whose matrix elements are real (see details in \textit{SI Appendix}). The search on both graphs is guaranteed to succeed within a bounded time, for any initial condition.  }    
    \label{fig1}
\end{figure}

 \begin{figure*}
    \centering
    \includegraphics[width=1.8\columnwidth]{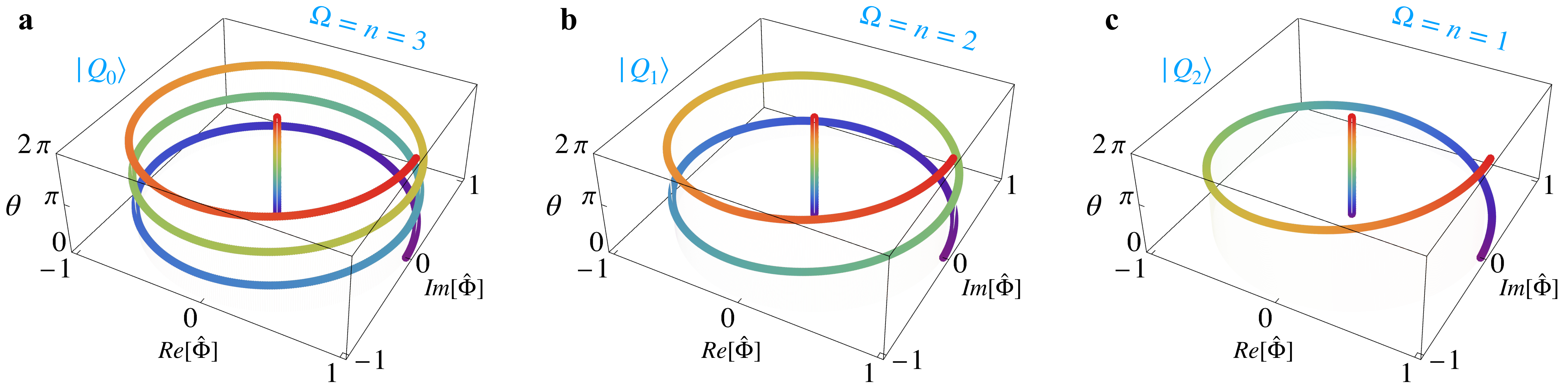}
    \caption{\textbf{Topological winding.} Plot of the generating function $\hat{\Phi}(\theta)$ versus $\theta$ for $N=3$. Here we choose the crawl Hamiltonian in Eq. (\ref{eq11}) and the generating function is given by Eq. (\ref{eq:generating function}). The initial states are the three $|Q_k\rangle$s which span the full Hilbert space. Due to the topology of the model, generating function forms a closed circle in the Laplace domain, hence the winding number is quantized as predicted in Eq. (\ref{eq:winding number of k}). As shown in the Figure, for the winding number of initial state $|Q_0\rangle$, we have $\Omega =3$ ({\bf a}). The winding number of $|Q_1\rangle$ is 2 ({\bf b}) and the winding number for $|Q_2\rangle$ is 1 ({\bf c}). These windings give the number of the measurements for the successful search. }    
    \label{fig:winding}
\end{figure*}

Let $F_n$ be the probability of detecting the system in state $\ket{\psi_{{\rm d}}}$ for the first time at $n\tau$. Then the total search probability of finding the quantum walker on the target state is $P_{{\rm det}}= \sum_{n=1} ^{\infty} F_n$. If $P_{{\rm det}}=1$  the mean search time is $\langle t \rangle = \tau \sum_{n=1} ^{\infty} n F_n$. The search probability $F_n$ is given in terms of the amplitudes $\phi_n$  of first detection, namely $F_n = |\phi_n|^2$ with \cite{Gruenbaum2013,Dhar2015,Friedman2017,Thiel2018,liu2021driving,Dhar2021}
\begin{equation}
\phi_n = \bra{\psi_{{\rm d}}} U(\tau) {\cal S}^{n-1}(\tau) \ket{\psi_{{\rm in}}},
\label{eq01}
\end{equation}
where the survival operator is ${\cal S}(\tau) = (1 - \ket{\psi_{{\rm d}}} \bra{\psi_{{\rm d}}} ) U(\tau)$, with $U(\tau)=\exp(-i H \tau)$ and $\hbar=1$. Here the back-action of the first $n-1$ repeated measurements is to repeatedly project out the amplitude of the target state $|\psi_{\rm d} \rangle$. In Eq. (\ref{eq01}) we have used the basic postulates of quantum theory with the projection $1 - \ket{\psi_{{\rm d}}} \bra{\psi_{{\rm d}}}$.

 As usual with these types of problems, the eigenvalues of the non-Hermitian operator ${\cal S}(\tau)$ are essential for the characterization of the process. The eigenvalues of ${\cal S}(\tau)$, denoted  $\xi$, are all on or inside the unit circle $|\xi|\le 1$, and the eigenvalues with $|\xi|=1$ correspond to dark states \cite{Krovi2006,darkstate,liu2021driving}. {\em Our goal} is to find $U(\tau)$ and the corresponding $H$ so that {\em all} the eigenvalues of ${\cal S}(\tau)$  are equal to zero. Intuitively, if all the eigenvalues are very small, the decay of $F_n$ is expected to be fast and the quantum search time will be minimized. It is also clear that if we find such an $H$, all the eigenvalues $\xi$ coalesce to the value $\xi=0$, meaning that we are engineering a method that yields a survival operator with a $N$-th order exceptional point, $N$ being the size of the system.

 The eigenvalues $\xi$ are given implicitly by
\begin{equation}
\mbox{det} | \xi- {\cal S} (\tau) | =
\xi \mbox{det} | \xi - U(\tau)| \bra{\psi_{{\rm d}}} \frac{1}{\xi - U(\tau)} \ket{\psi_{{\rm d}}} = 0
\label{eq02}
\end{equation}
where we have used the matrix determinant lemma, see \textit{Materials and Methods}. Clearly, the system always has at least one solution $\xi=0$. Let $H \ket{E_k} = E_k \ket{E_k}$ where $k=0, \cdots, N-1$ and as usual we may expand $\ket{\psi_{{\rm d}}}= \sum_{k=0} ^{N-1} \bra{E_k} \psi_{{\rm d}} \rangle \ket{ E_k}$, and then 
\begin{equation}
\bra{\psi_{{\rm d}}} \frac{1}{\xi - U(\tau)}  \ket{\psi_{{\rm d}}} =  \sum_{k=0} ^{N-1} 
 { p_k \over \xi - \exp( - i E_k \tau) } 
\label{eq03}
\end{equation}
with $\mbox{det} | \xi - U(\tau)| = \prod_{k=0} ^{N-1} [ \xi - \exp(- i E_k \tau)]$. Here $p_k= |\bra{E_k}  \psi_{{\rm d}}\rangle|^2$ is the square of the overlap between the  energy state $\ket{E_k}$ and the detected state. Our first requirement is that the system is such that $p_k \ne 0$ for all the energy states $\ket{E_k}$, and that there is no degeneracy i.e., $\exp( - i E_k \tau) \neq \exp( - i E_m \tau)$ for any choice of $m \neq k$. Physically this demand means that we exclude dark states so that $|\xi|<1$ and hence the eigenvalues  satisfy $\mbox{det} | \xi - U(\tau)| \neq 0$. Using Eqs. (\ref{eq02},\ref{eq03}) it is not difficult to show that the eigenvalue problem reduces to finding the solution of 
\begin{equation}
\xi \sum_{k=0} ^{N -1} { p_k \over \xi - \exp( - i E_k \tau) }= 0.
\label{eq04}
\end{equation}
We now engineer the system in such a way that the only solution is the degenerate solution with $\xi=0$. As we will shortly show, the following requirement is sufficient 
\begin{equation}
p_k = { 1 \over N} \ \ \mbox{and} \ \  E_k \tau = { 2 \pi k \over N} .
\label{eq05}
\end{equation}
We see that the energy levels are equally spaced, which intuitively is expected as this causes the periodicities in the dynamics to resonate at specific times, enhancing constructive interference. More specifically, we will soon choose $E_k= \gamma k$ where $\gamma$ has units of energy, and then $\tau=2 \pi/\Delta E$ where $\Delta E=E_{max} -E_{min}$ is the energy gap between the ground and largest energy in the spectrum. We note here that a relativistic mass-less free particle, with energy  $E= \sqrt{ m^2 C^4 + c^2 p^2}$, and $m=0$, has a dispersion $E_k \propto p \propto k$, instead of the well-known Schr\"{o}dinger  dispersion of a free particle  $E_k \sim k^2$. Hence the energy spectrum we find in Eq. (\ref{eq05}) is essentially relativistic; the consequence of this for search will be discussed later. We also see that the overlaps  $p_k$  are $k$-independent.  To verify these requirements, insert Eq. (\ref{eq05}) in Eq. (\ref{eq04}) and then with summation formulas (see \textit{SI Appendix}) we have
\begin{equation}
{ \xi \over N} \sum_{k=0} ^{N -1} { 1 \over \xi - \exp(-i 2 \pi k/N) } =
- {\xi^N \over 1 - \xi^N}=0
\label{eq06}
\end{equation}
and the only possible solution is $\xi=0$. 
We see that for a  quantum system satisfying Eq.
(\ref{eq05}),
the survival operator has a $N$-fold degenerate eigenvalues at
$\xi=0$, as we aimed for. The order of the exceptional point is equal to the size of the Hilbert space $N$, namely 
\begin{equation}
	\xi=0, \quad \quad  N\text{-th order exceptional point}.
\end{equation}
It can also be shown that Eq. (\ref{eq05}) is in fact a necessary condition for a degree $N$ exceptional point (see \textit{SI Appendix}).
Further, all the right and left eigenvectors also coalesce with $|\xi^R\rangle = U(\tau)^{-1} |\psi_d\rangle = U(-\tau) |\psi_d\rangle $ and $\langle \xi^L|=\langle \psi_{\rm d}|$.  Before constructing the graph that yields
this result we study its general consequences for search.

\section*{Efficient quantum search and quantized topological winding number}

 We denote $H_s$, $U_s$ and ${\cal S}_s$  the Hamiltonian, unitary and survival operator
for a system  that satisfies the efficient search conditions 
Eq. (\ref{eq05}) 
and in this notation we omit the dependence on $\tau$.
We define the states $\ket{ Q_k} = (U_s)^k \ket{\psi_{\rm d}}$ with $k=0, \cdots,
N-1$. The operators $U_s$ and ${\cal S}_s$ acting on these
 states give
\begin{equation}
\begin{array}{l l}
U_s \ket{Q_{N-1}} =  \ket{\psi_{{\rm d}}}, \  & U_s \ket{Q_{k}} = \ket{Q_{k+1}} \  \mbox{if}  \ k\neq N-1, \\
{\cal S}_s \ket{Q_{N-1}} = 0, \   & {\cal S}_s \ket{Q_{k}} = \ket{Q_{k+1}} \ \mbox{if} \ k\neq N-1.
\end{array}
\label{eq07}
\end{equation}
These formulas are mathematically straightforward,  for example $U_s \ket{Q_{N-1}} = (U_s)^N \ket{\psi_{\rm d}}
=\sum_{k=0} ^{N-1} (U_s)^N \bra{E_k} \psi_{{\rm d}} \rangle \ket{E_k}=
\sum_{k=0} ^{N-1} \exp( - i E_k \tau N)   
\bra{E_k} \psi_{{\rm d}} \rangle \ket{E_k}=\ket{\psi_{{\rm d}}} = \ket{Q_0}$ where we used
Eq. (\ref{eq05}) and hence $\exp( - i E_k \tau N) = 1$. 
We see that both ${\cal S}_s$  and $U_s$ are shift operators, their difference being the action on the boundary term $\ket{Q_{N-1}}$. We can also show that
the states $\ket{Q_k}$ are orthonormal $\bra{Q_l} Q_m \rangle = \delta_{lm}$ (see \textit{SI Appendix}) and they
form a complete set spanning any initial condition in the Hilbert space. From here we reach
the following conclusions. First, consider an initial condition which is a $\ket{Q_k}$ state,
then following Eq. (\ref{eq01}) we consider the operation $({\cal S}_s)^{n} \ket{Q_k}$ and
using Eq. (\ref{eq07}) we obtain
\begin{equation}
\phi_n = 
\left\{ 
\begin{array}{l l}
1 \ \ & \ \mbox{if} \ \ n=N-k , \\
0 \ \ & \ \mbox{otherwise} .
\end{array}
\right. 
\label{eq08}
\end{equation}
This means that we detect the target with probability one at time $(N-k)\tau$, hence
the detection process is deterministic as the fluctuations of the detection time vanish.
Then when $|\psi_0\rangle = |Q_k\rangle$, we have
\begin{equation}
 P_{{\rm det}}=1, \quad \langle t \rangle = t = \tau(N-k), \quad  \text{Var}(t) = 0. 
\end{equation}
For a more general initial condition, exploiting the fact that the states $\ket{Q_k}$ form
a complete set and the linearity of Eq. (\ref{eq01}) with respect to the initial
condition, the probability of first detection  $F_n=|\phi_n|^2$ is
\begin{equation}
F_n = 
\left\{ 
\begin{array}{l l}
|\bra{Q_{N-n}} \psi_{{\rm in}}\rangle |^2   \ \ & \ \mbox{for} \ \ n=1, \cdots, N, \\
0 \ \ & \ \mbox{otherwise} .
\end{array}
\right. 
\label{eq09}
\end{equation}
This implies a guaranteed search, since even in the absence of knowledge about the initial
condition, the search will find the target with at most $N$ operations.
From here it also follows that we have an upper bound on the search time for any initial state
\begin{equation}
	 t \le \tau N = \frac{2 \pi k}{E_k} = \frac{2\pi}{\gamma}.
\end{equation}
This upper bound is $N$ independent, so the maximum search time does not increase with the system size. The upper limit is found when the initial condition is the target state $\ket{\psi_{{\rm in}}} = \ket{\psi_{{\rm d}}}$. To conclude, for a quantum walker starting from an unknown initial state, i.e., a black-box problem, our strategy will find this walker at the target state within a fixed time with probability one.

We next explore the topological properties of the efficient quantum search. In spatially periodic systems, such as the topological materials, their topologies are revealed by the Chern number or winding number of the Bloch Hamiltonian in the band-theory framework \cite{RevModPhys.88.035005}. For our model, the periodicity originates from the stroboscopic measurements. Hence instead of a Brillouin zone in $k$ space, here we investigate the topology of the system in the Laplace domain \cite{Friedman2017}. Using the $Z$ transform, the generating function of the search amplitude $\phi_n$ reads:  
 \begin{equation}
 	\hat{\Phi}(\theta)=\sum_{n=1}^{\infty} e^{i n \theta} \phi_n = \frac{\langle \psi_{\rm d}|\hat{{\cal U}}(\theta)|\psi_{\rm 0}\rangle }{1+ \langle \psi_{\rm d}|\hat{{\cal U}}( \theta )|\psi_{\rm d}\rangle},
 	\label{eq:generating function}
 \end{equation}
  where $\hat{{\cal U}}( \theta )= \sum_{n=1}^{\infty} e^{i n \theta} U(n\tau) =e^{i \theta} U(\tau)/(1- e^{i \theta} U(\tau))$ is the generating function of $U(n\tau)$ \cite{Thiel2018}. The statistics of the search process can be calculated in terms of the generating function. For example, the total search probability $P_{det} = 1/(2\pi)  \int_{0}^{2\pi}  d\theta | \hat{\Phi}( \theta ) |^2 $ and the mean search time $\langle t \rangle  = \tau/(2\pi i) \int_{0}^{2 \pi } d \theta [\hat{\Phi}\left( \theta \right)]^{*}[ \partial_{\theta} \hat{\Phi}\left( \theta \right)],$ where $*$ is the complex conjugate of the generating function \cite{Friedman2017}.
  
 With Eq. (\ref{eq:generating function}), we calculate the winding number when the quantum system meets the conditions in Eq. (\ref{eq05}), corresponding to an $N$-th order exceptional point.  The winding number is quantized and characterized by the choice of the initial state. When $|\psi_0\rangle = |Q_k\rangle$, the winding number  $\Omega$ reads
  \begin{equation}
 	  \Omega = \frac{1}{2 \pi i} \int_{0}^{2\pi}   d \theta \ \partial_{\theta} \ln[ \hat{\Phi}(\theta)] =N-k.
 	  \label{eq:winding number of k}
 \end{equation} 
 Using Eq. (\ref{eq09}), this quantized winding number equals the number of measurement attempts needed to detect the walker with probability unity. It is in this sense that the search process is related to the topology of the model in Laplace space and the search times for states $|Q_k\rangle$ are given in terms of $\tau$ multiplied by the number of windings, i.e., $t= \Omega \tau$. We plot $\hat{\Phi}(\theta)$ in Fig. \ref{fig:winding} for $N=3$. Note here we use the crawl Hamiltonian later derived in Eq. (\ref{eq11}) for illustration.  As shown in the Figure, the $\hat{\Phi}(\theta)$ forms closed circles and the number of times it rotates around the center is equal to $\Omega$. 
  
  \begin{figure*}
    \centering
    \includegraphics[width=1.4\columnwidth]{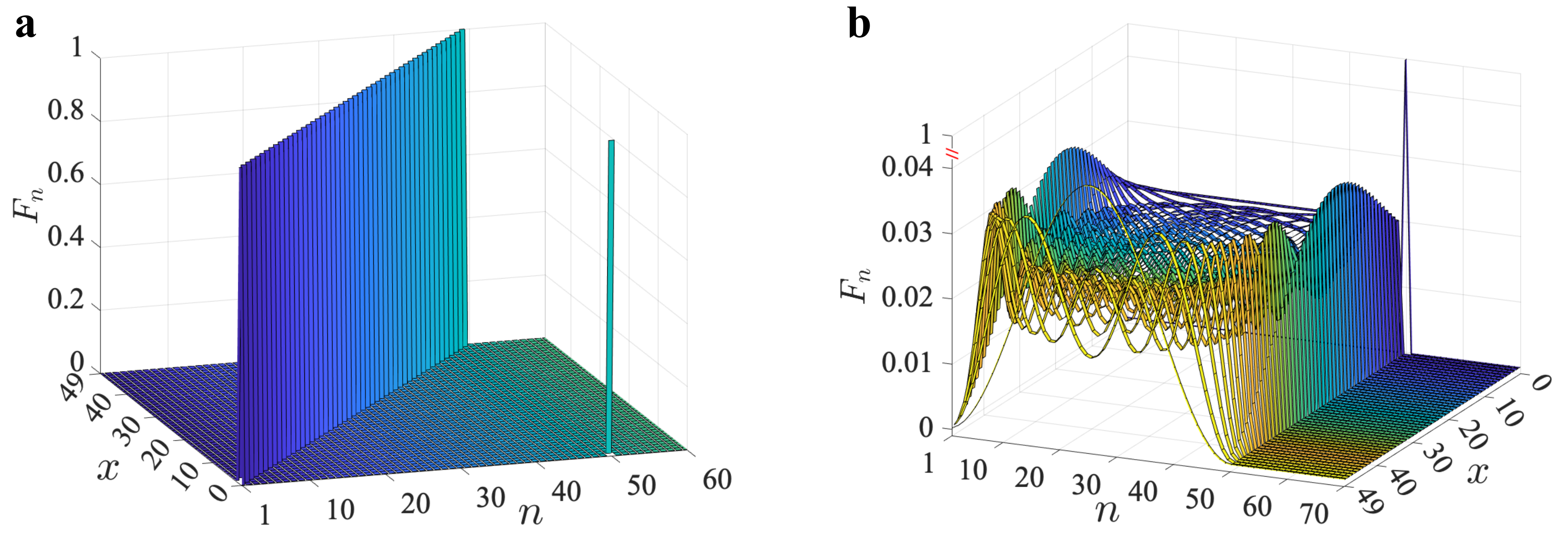}
    \caption{\textbf{Guaranteed and fast quantum search.} Detection probability $F_n$ versus $n$ for the crawl ({\bf a}) and funnel ({\bf b}) model.  Here the graph has $N=50$ nodes and we present results with initial states localized on one of the nodes $\ket{\psi_{{\rm in}}}=\ket{x}$, $x=0, 1, \cdots $ and $|\psi_{{\rm d}}\rangle=|0\rangle$. For the crawl search we find a deterministic outcome of the process, where $F_n=1$ when  $n= x$ (for $x=0$, $F_{50}=1$).  For the funnel model ({\bf b}) notice the sharp cutoff of $F_n$ for $n>N=50$. For any initial condition, the detection of the state is guaranteed with probability one, within at most $N$ measurements, which we call guaranteed search. In ({\bf b}) notice the peak of height one for $n=50$,  when the initial condition is the same as the detected state.  The upper bound for the search time is $t_{max}= \tau N= 2\pi$.   }    
    \label{fig2}
\end{figure*}

\section*{Examples of designed quantum graphs: crawl and funnel models}

 What are the tight-binding Hamiltonians of size $N \times N$, that yield a guaranteed search? The condition Eq. (\ref{eq05}) admits many types of solutions, and here we present two that have certain advantages. 
 \subsection*{Crawl Model} 
 
 First, we present an approach where the nodes of the graph are the states $\ket{Q_k}$. This is clearly useful since this means that we can start the process with the wave packet on one node of the graph and find the walker with probability one after a fixed time at any other node, which we call deterministic search, as the fluctuations vanish. We use $H=\sum_{k} E_k \ket{E_k} \bra{E_k}$ and for the equal distance energies we set $E_0= 0, E_1= \gamma, \cdots, E_{N-1} = (N-1)\gamma$ and Eq. (\ref{eq05}) gives $\tau= 2 \pi/N\gamma$. More generally $\tau= 2 \pi/ N \gamma + 2 j \pi/\gamma $ and $j$ is a non-negative integer. In this system the states $\ket{x}$ with $x=0,1, \cdots, N-1$ are the nodes of the graph, see Fig. \ref{fig1}.  To perform this trick let
\begin{equation}
\ket{E_k} = \left\{ 1 , e^{ i \theta_k}, e^{ i 2  \theta_k} , \cdots , e^{i(N-1) \theta_k} \right\}^{T}/\sqrt{N}
\label{eq10}
\end{equation}
where $\theta_k = 2 \pi k/N$. This eigenstate is a discrete Fourier wave, which is related to the ``relativistic" linear dispersion in Eq. (\ref{eq05}), and Dirac physics as discussed below. Clearly Eq. (\ref{eq10}) gives  $p_k = 1/N$ and hence
\begin{equation}
\begin{array}{c}
H_{{\rm Crawl}} =
 \ \\
\gamma\left[
\begin{array}{c c c c c}
0 &  {1 \over 1 - e^{i \theta_1}} &  {1 \over 1 - e^{i \theta_2}} & \cdots &   {1 \over 1 - e^{i \theta_{N-1}}}  \\
 {1 \over 1 - e^{-i \theta_1}} & 0 &   {1 \over 1 - e^{i \theta_1 }} & \cdots &   {1 \over 1 - e^{i \theta_{N-2}}}  \\
 {1 \over 1 - e^{-i \theta_2}} &  {1 \over 1 - e^{-i \theta_1}} & 0 & \cdots & {1 \over 1 - e^{i \theta_{N-3}}}  \\ 
 \vdots & \vdots  & \vdots  & \ddots  & \vdots \\ 
 {1 \over 1 - e^{-i \theta_{N-1}}} &    {1 \over 1 - e^{-i\theta_{N-2}}} &   {1 \over 1 - e^{-i \theta_{N-3}}} & \cdots & 0 
\end{array}
\right]
\end{array}
\label{eq11}
\end{equation}
which we call the Crawl Hamiltonian, see a schematic diagram in Fig. \ref{fig1}({\bf a}). This system, as discussed below, breaks time-reversal symmetry. Namely, the uni-directional movement of the packet can be reversed by changing $H_{{\rm Crawl}}$ to its complex $H_{{\rm Crawl}}^*$, which is a feature of time-reversal.  In Fig. \ref{fig2}({\bf a}) we plot $F_n$ for a system with $N=50$ and where the target state is $\ket{\psi_{{\rm d}}}= \ket{0}$. We choose local initial conditions such that $\ket{\psi_{{\rm in}}} = \ket{x}$, hence we are considering a transition from $x$ to $0$ and in the plot we choose $x=0,1, \cdots, 49$. We see that $F_n$ is sharply peaked and is equal to unity when $n= x$ [Fig. \ref{fig2}({\bf a})]. This type of deterministic search is not found for classical random walks, and relies on the fact that the quantum wave packet can, at specific times of the evolution, be localized on a single node, while at prior measurement times the wave packet vanishes on the node. This $H_{{\rm Crawl}}$ is very much related to unitary state transfer in the absence of measurements, as discussed below.

The efficient quantum walks we found here are reminiscent of the physics of a massless Dirac particle in dimension one.  First, the energy is linear in $k$, Eq. (\ref{eq05}). Second, in the crawl model the energy states are discrete free waves, and finally, due to time reversal breaking, the wave packet can travel either clockwise or anti-clockwise, somewhat similar to a particle and anti-particle. But why do we find this relation between our problem and these relativistic effects? We started this work with the demand that the eigenvalues of ${\cal S}$, are real and all coalesce to zero, to speed up the search process. We then added rotational invariance of the searching process, such that all nodes of the graph are identical, namely no matter what the detected state, $p_k=1/N$ on every node of the specially designed graph. We then naturally find the ideal search for a quasi-particle with no dispersion, at least at the measurement times. Namely, a wave packet that is widening will create a less efficient search, in the sense that it renders impossible the absolute detection of the particles in a single measurement made at a node, a feature that is also revealed by the quantized winding number in the Laplace domain. Similarly inspired by a massless Dirac particle, consider the trivial wave  equation in continuous space and time in dimension one, $\partial_x \psi(x,t) = \partial_t \psi(x,t)$, the solution is $\psi(x,t) = \int g(k) \exp( i (k x - w_k t))$ and $g(k)$ is the initial  packet in momentum space. For a localized initial condition and using $w=k$, we get a delta traveling wave, in close analogy with what we find in discrete space. Of course, the underlying dynamics in our case are controlled by the Schr\"{o}dinger equation, but the Hamiltonian under study yields effective motion in which space and time are treated on the same footing.  Finally, Dirac's wave function in dimension one has two components. Similarly, we have a  particle traveling clockwise and anti-clockwise, in fact, at least in principle, we can switch between these modes, if in the middle of the experiment we replace the Hermitian crawl $H$ with its complex conjugate.
 
\subsection*{Funnel Model} 

An alternative approach that uses on-site energies to direct
the search to a specific node denoted $\ket{\psi_{\rm d}}= \ket{0}$ is now considered.
Here the process does not break time-reversal symmetry. 
As before, the spatial nodes of the graph are denoted $|x_i\rangle$, and $i=0,1, \cdots, N-1$. We still have condition to fulfill Eq. (\ref{eq05})  and
we start with the normalised state  $\ket{E_0} = \left\{ 1/\sqrt{N}, -\sqrt{ N -1}/\sqrt{N}, 0, \cdots \right\}^T$
in agreement with the second condition in Eq. (\ref{eq05}). 
The next energy state is constructed such that it is normalized and orthogonal
to the first one and has an overlap $1/N$ with the detected state,
$\ket{E_1} = \left\{1/\sqrt{N},  1 /\sqrt{ N (N-1)}, -\sqrt{N-2}/\sqrt{N-1}. 0, 0, \cdots\right\}^T$.
The process of constructing these states is then continued (\textit{SI Appendix}), and exploiting the demand that energy levels be equidistant,
Eq. (\ref{eq05}), we find

\begin{equation}
\begin{array}{c}
H_{{\rm Funnel}} =
\frac{\gamma}{2} \left[
\begin{array}{c c c c c c}
N-1 & \sqrt{N-1} &   \cdots &  \sqrt{\frac{2}{N}}  \\
\sqrt{N-1}  & 1 &    \cdots  & \sqrt{\frac{2}{N^2-N}}  \\

 \vdots & \vdots   & \ddots  & \vdots \\ 
\sqrt{\frac{2}{N}} & \sqrt{\frac{2}{N^2-N}} &   \cdots & 2N -3 
\end{array}
\right]
\end{array}
\label{eq:funnel}
\end{equation}
with matrix elements $	H_{0,m}=\sqrt{(N-m)(N-m+1)/N}$ ($m\neq 0$) and $  H_{j, m } =  \sqrt{(N-m)(N-m+1)/[(N-j+1)(N-j)]}$ ($j \neq 0, m$). We call this approach the funnel model.
This type of system is schematically shown in Fig. \ref{fig1}({\bf b}) while in Fig. \ref{fig2}({\bf b})
we present the detection probabilities for localized initial conditions.
Fig. \ref{fig2}({\bf b}) illustrates a sharp cutoff, namely $F_n=0$ for any $n>N$  and
thus the search is guaranteed to succeed in a finite time, a feature completely absent in 
classical random walks or quantum walks on non-specialized graphs. Interestingly, if the initial state is the same as the detected one, corresponding to what is known as the return problem \cite{Gruenbaum2013}, $F_{50}=1$, otherwise it is zero. This means the system is detected exactly after $N$ attempts, and this feature is universal, namely for any search $H_s$, satisfying Eq. (\ref{eq05}), and for $|\psi_{\rm d} \rangle  = |\psi_0 \rangle$, $F_N =1$ and $F_{n \neq N} =0$. We will discuss this surprising effect in more detail in the discussion.

\begin{figure}
    \centering
    \includegraphics[width=0.91\columnwidth]{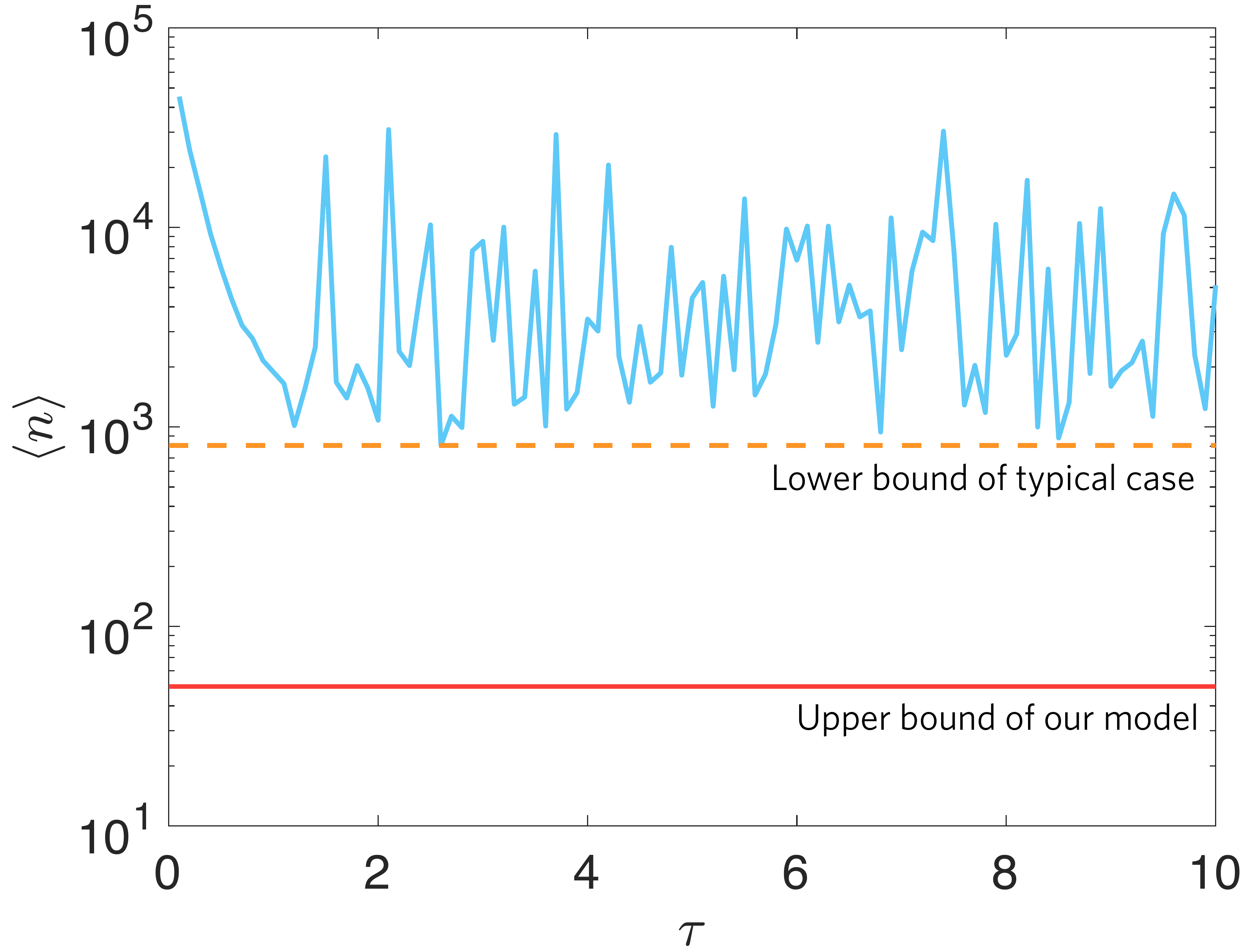}
    \caption{\textbf{Comparison with typical graph.} We plot the mean measurement numbers for a full randomly connected SK graph for different choices of $\tau$ with stroboscopic search protocol. Here we choose $N=50$ and for the designed graph the upper bound of the measurement number is $50$ with time $t \leq 2\pi/\gamma$ for $\tau = 2\pi/\Delta E $. The search on the typical graph is much slower than that. As shown in the Figure, the lower bound of the SK graph is much bigger that the upper bound of our models. It is clear the designed graphs boost the quantum search process. }    
    \label{fig:comparison with random H}
\end{figure}

 \begin{figure}
    \centering
    \includegraphics[width=0.9\columnwidth]{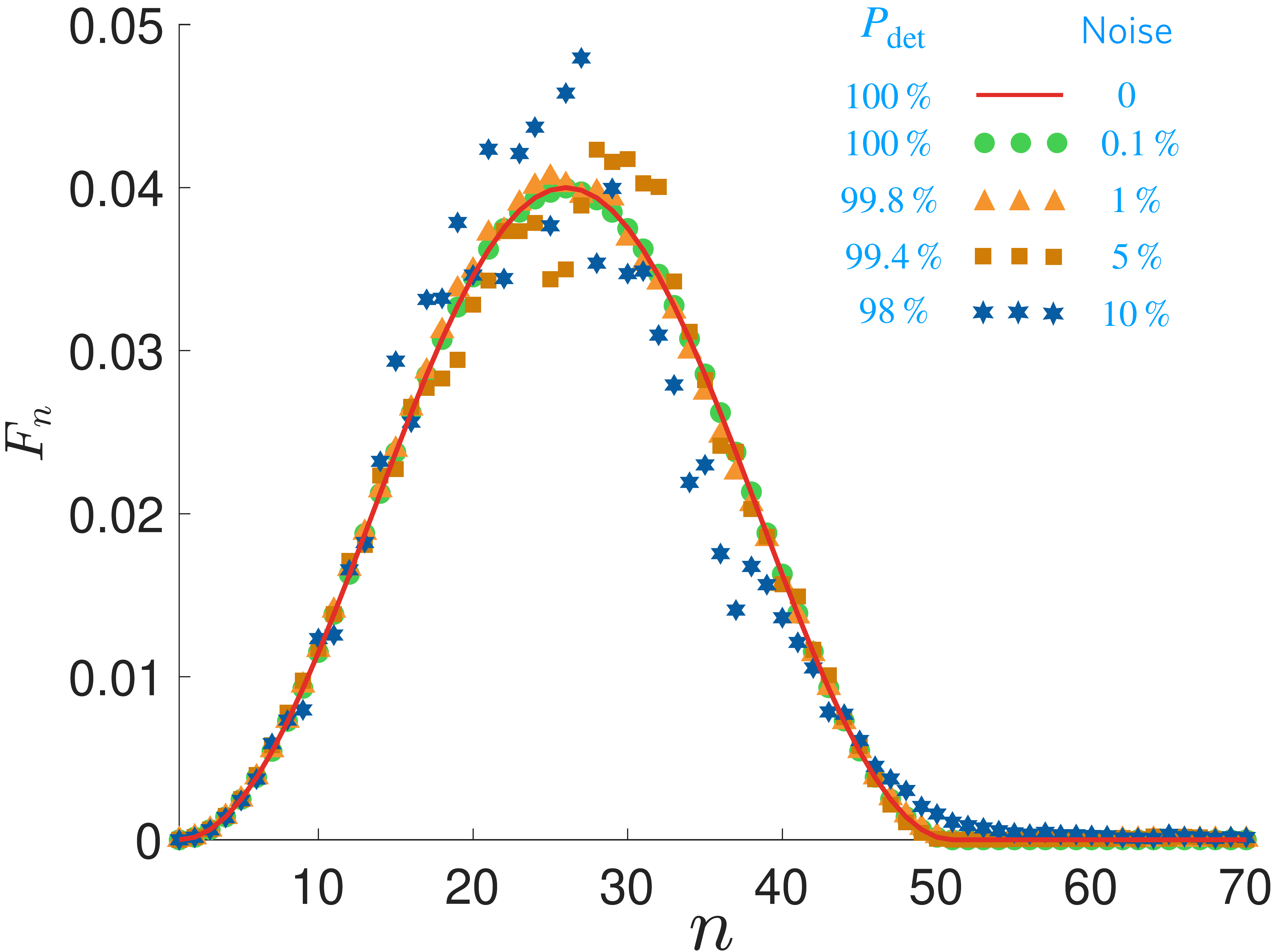}
    \caption{ \textbf{Noise robustness.} We plot the search probability $F_n$ of the $50$-site funnel model with the initial state being a node of the graph. We introduce random noise to $\tau$ with different magnitudes from $0.1\%$ (green dots) to $10\%$ (blue hexagrams). As shown in the Figure, the $F_n$ is robust to the noise and the search probability is nearly zero for $n > 50$. We calculate the total search probability $P_{\rm det}$ within the first 50 measurements for 1000 realizations.  The guaranteed search reminds even for comparable large noise, where the $P_{\rm det} \sim 98\%$ with noise $10\%$.}    
    \label{fig:noise tau}
\end{figure}

\section*{Comparison to the SK graph, influence of noise, and connections to quantum state transfer}

 As a comparison, we calculate the search time for finding the quantum walker on a full randomly connected Sherrington-Kirkpatrick (SK) model \cite{Harrigan2021} with $50$ nodes, in the sense that the designed coupling in our model is replaced by the random connections. As we have shown, for our designed graphs with $50$ nodes, the walker will be detected within $50$ measurement attempts and the upper bound of the search time is $2\pi/\gamma$. As shown in Fig. \ref{fig:comparison with random H}, on the SK model, the mean number of the measurement to find the walker is much larger than our designed graphs ($\langle n \rangle \sim 10^3 \gg 50$) and fluctuates strongly for different choices of $\tau$. When $\tau$ is small, the detection times diverge due to the quantum Zeno effect \cite{Misra1997,Itano1990,Facchi2001inv,PhysRevA.103.032221}.

We also plot the cases when there is noise on $\tau$ in Fig. \ref{fig:noise tau}. Here we choose the funnel model with $N=50$ and the initial state is $|\psi_{\rm 0}\rangle=|49\rangle$. For each sampling interval $\tau$, we use the designed $\tau$ [Eq. (\ref{eq05})] with the noise generated from a uniform distribution, see \textit{Materials and Methods}. We plot the detection probability versus measurements for different noise strengths. The model shows robustness to the effects of the noise, where the search is still guaranteed to succeed, namely $\sum_{n=1}^{50} F_n \sim 1$.  The detection probability is close to zero for $n > 50$ as shown in Fig. \ref{fig:noise tau}. As a more realistic test of the guaranteed search, we calculate the search probability $P_{\rm det}$ within 50 measurements for $1000$ realizations. We find $P_{\rm det} \sim 0.98$ even under $10\%$ noise, hence the approach is robust.

\begin{figure}
    \centering
    \includegraphics[width=\columnwidth]{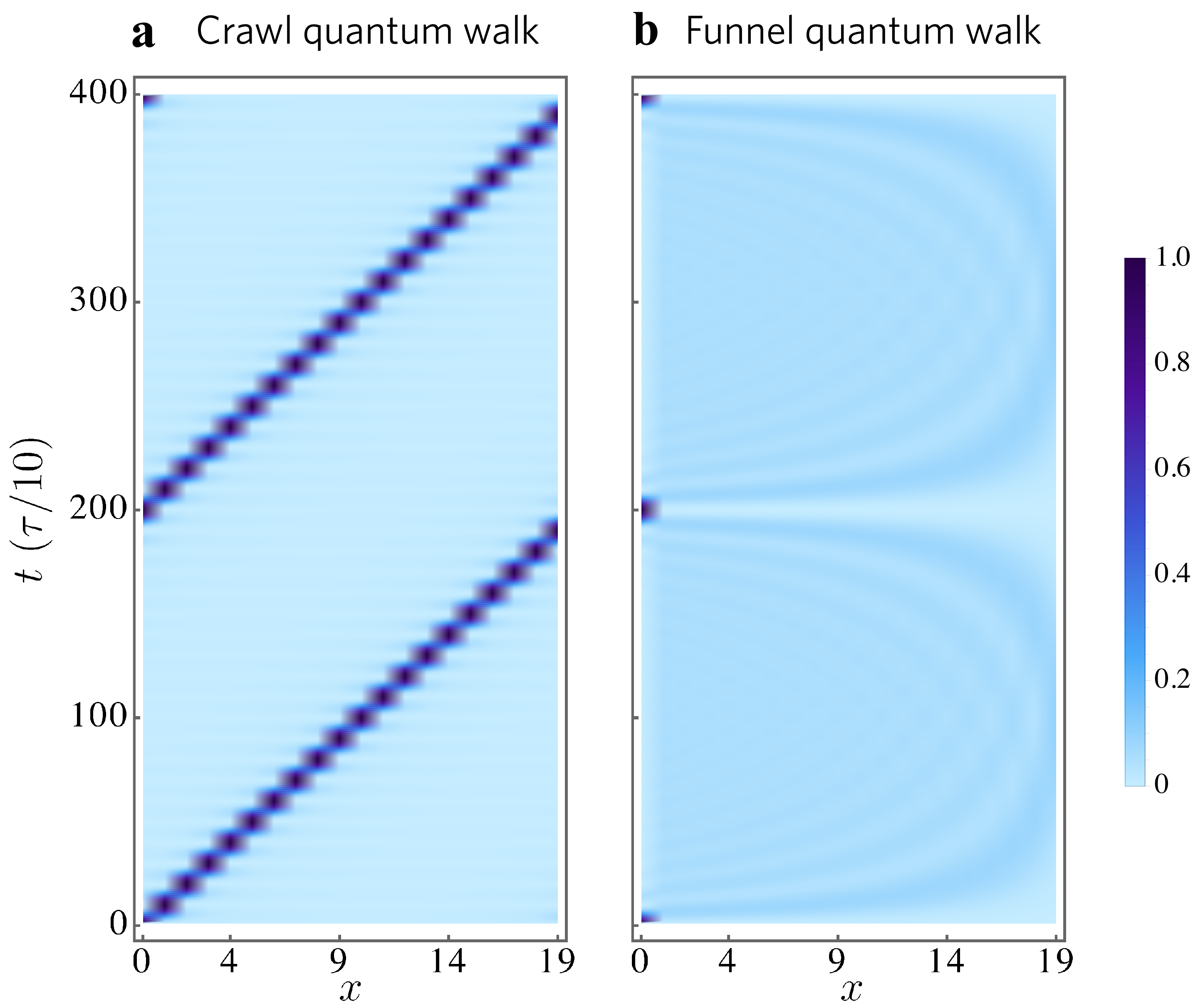}
    \caption{\textbf{Non-monitored quantum walk and perfect quantum state transfer.} The time evolution of the wave packet on the crawl ({\bf a}) and funnel ({\bf b}) graphs for non-monitored quantum walks. The color code gives the probability of finding the walker on a node.  For the crawl graph, the wave function is fully localized one node after the other at times $\tau, 2\tau, \cdots$, a feature which is vital for state transfer to any node in the system.  Both quantum walks exhibit revival, namely the wave function returns to its initial state, at time $ N \tau $. }    
    \label{fig3}
\end{figure}

We investigate the effects of repeated measurements on the search for an unknown initial state by designing special graphs. In some special cases, the measurement does not interact with the unitary dynamics. To see this special case, consider a walker that starts from a node of the graph, in other words, we now assume the complete knowledge of the initial state, which is a special localized state. As mentioned, using the crawl model Eq. (\ref{eq11}) the unitary $U(\tau)$ is the shift operator, which shifts the particle from one node to the other (see Fig. \ref{fig3}). Throughout the evolution, the wave function $|\psi(t)\rangle$ is zero on $|\psi_{\rm d}\rangle$. So, the measurements do not interfere with the walker. Only after a number of shifts (depending on the distance between nodes), the shift operations transport the system to the target state, and the measurement records the system with probability 1. Namely, if we start on a node $\ket{\psi_0}= \ket{x_0}$ and focus on the final node $\ket{\psi_{\rm d}}= \ket{x_f}$ the (non-monitored)  success probability is
\begin{equation}
| \langle \psi(t) | \psi_{{\rm d}} \rangle|^2 = 1
\end{equation} 
at times $t = (x_f-x_0)\tau$. Therefore, since measurements do not destroy the unitarity till the final measurement, our result is equivalent to quantum state transfer with unitary dynamics for this special case, which has been considered by Kostak {\em et al.} \cite{Jex2007}. For comparison for the funnel model, even if the system starts from a special localized node, we see that the wave function spreads to all the nodes of the graph. So the measurements will interact and disturb the evolution of the funnel walker during the whole search process, until it is caught by the detector.

\section*{Discussion}

We have designed a survival operator ${\cal S}(\tau)$, with an exceptional point whose degeneracy is the size of the Hilbert space. Such an exceptional point reaches the highest order of degeneracy possible in the model and can be designed as large as possible. This is certainly an advance in exceptional physics compared to previous results considering second or third order exceptional points. In general, for an $N$-dimensional system, to find the highest order exceptional point, one needs to solve an $N$-th order characteristic polynomial \cite{PhysRevLett.123.213901,Hodaei2017}, which, in principle, is difficult when the system is large. Here we show that the high order exceptional point can be designed by exploiting the symmetry of the model, which leads to the two conditions we discussed in Eq. (\ref{eq05}). At the exceptional point, the vector space is severely skewed, as all the eigenstates of ${\cal S}(\tau)$ coincide. Obviously, this means that the single eigenstate of ${\cal S}(\tau)$ cannot be used to construct a full basis.   So can we find a new basis that spans the Hilbert space, which should also be connected to the exceptional properties of the model?  This challenge is solved by the states $|Q_k\rangle$ we proposed, which form a full basis that can be used to expand any initial state of the walker, and is determined by the system parameters at the exceptional point. In this new basis, the survival operator ${\cal S}(\tau)$ becomes a shift operator Eq. (\ref{eq07}). Roughly speaking, this new basis can be considered to play the role of the energy eigenbasis found for unitary dynamics. The new basis $|Q_k \rangle$ is an efficient tool for studying the quantum search process. We find a full basis using the exceptional point, which is related to the fact that the degree of the exceptional point here is the size of the Hilbert space, while if the degree of the exceptional point is less than that, the effect would not have been found.

Remarkably, the search probability is sharply peaked, i.e., $F_{n=N} =1$, for the case $|\psi_{\rm d}\rangle = |\psi_{\rm in}\rangle$, as shown in Fig. \ref{fig2}({\bf b}). Since the initial state here is the same as the target one, this is called in the literature the return problem \cite{Gruenbaum2013}. This is a generic property for all the designed graphs and describes the special recurrence property in repeatedly monitored quantum walks. To see this note that choosing $k=0$, $|Q_0 \rangle = |\psi_{\rm d} \rangle$, and then use Eq. (\ref{eq10}). Physically, this shows the wave function always has destructive interference on the search target at times $n \tau$ with $n=1, \cdots, N-1$ and fully constructive interference, which collects all the amplitude of wave function, at time $N\tau$ at $|\psi_{\rm d}\rangle$.  In connection with previous results, Gr\"{u}nbaum \textit{et al.} have shown that the mean search time for the return problem is quantized and equals the effective dimension of the system \cite{Gruenbaum2013,Bourgain2014}, which is related to the topology of the Schur functions. This result in our case means that the average $ \langle n \rangle = N$ for the return problem. However, as shown in our work, for purposely designed graphs, we have $F_n= 0$ for $n > N$ for the guaranteed search. These two results together mean that in our case we must have $F_N=1$ for the return problem. The result is therefore the sharp peak seen in Fig. \ref{fig2}.  This mean that in general for the return problem $\text{Var}(n) \neq 0$, while we have $\text{ Var}(n) = 0$, namely no fluctuations at all of the return time. This indicates a specialized recurrence on our designed graphs, which is absent in previous results.      

 The resulting special purpose family of Hamiltonians allows for guaranteed search.  The main condition Eq. (\ref{eq05}) still allows for further freedom in the design of the search process. For specialized states $\ket{Q_i}$, which are used to expand the Hilbert space as we discussed, the monitored search process is deterministic as the fluctuations in the detection attempt vanish. Our work shows a connection between guaranteed search and the topology, see Fig. \ref{fig:winding}.  Very generally, starting with state $|Q_i \rangle$ the generating function winds, the number of windings gives the number of measurements till detection. Hence the search time can also be expressed in terms of the winding number times $\tau$, i.e., $ t= \Omega \tau$.  For a random unknown initial state, the quantum walks we designed here are guaranteed to succeed in a bounded time. Usually, topology is related to some protected physical reality that is insensitive to sources of noise. Also in our case, the topology is related to a physically robust result, i.e. protecting the search in the sense that it is secured to be detected, up until a fixed time, no matter what is the initial state.

 For the crawl model, the search is effectively uni-directional in a system that conserves energy. The Hamiltonian is independent of the choice of the target state, and in that sense, the system exhibits universal search. The $H_{{\rm Crawl}}$ breaks time-reversal symmetry, and is related to the Dirac equation. In contrast, the funnel model does not break time-reversal symmetry, but the target state is unique.

 To conclude, our search is globally optimal with respect to any other search algorithm, in the sense that the success probability is unity within a finite time even for large systems and for all the initial conditions.
 
 \section{Methods}

\subsection*{Matrix determinant lemma}
We provide details on the derivation of Eq. (\ref{eq02}) using the matrix determinant lemma. Suppose $A$ is an invertible square matrix and $u$, $v$ are column vectors, then the matrix determinant lemma states:
\begin{equation}
	\text{det}|A+uv^T|=(1+v^T A^{-1} u)\text{det}|A|,
	\label{smeq01}
\end{equation}
where $uv^T$ is the outer product of the vectors $u$ and $v$. We are interested in the eigenvalues of the survival operator $\text{det}|\xi- {\cal S}(\tau) |=0$ with  ${\cal S}(\tau) = (1-|\psi_{\rm d}\rangle \langle \psi_{\rm d}|)U(\tau)$  as defined in the main text. Substituting ${\cal S}$ into the matrix determinant, we have
\begin{equation}
	\text{det}|\xi-{\cal S}(\tau)|=\text{det}|\xi-U(\tau)+|\psi_{\rm d}\rangle \langle \psi_{\rm d}| U(\tau)|.
\end{equation}
As discussed in the main text, to prevent the appearance of dark states, which are not optimal for the quantum search, we have conditioned $\text{det}|\xi -U(\tau)|\neq 0$. Hence $\xi-U(\tau)$ is an invertible square matrix. If we denote $\xi-U(\tau)$ as the matrix $A$, it fits the condition for the matrix determinant lemma. We then let $u= |\psi_{\rm d}\rangle$, and $v^T= \langle \psi_{\rm d}| U(\tau)$. Using Eq. (\ref{smeq01}), we have get Eq. (\ref{eq02}) used in the main text.

\subsection*{Numerical Simulation Approach}

To prepare the plots, we simulate the search process directly based on Eq. (\ref{eq01}). We first construct the search Hamiltonians for the crawl graph and funnel graph using Eqs. (\ref{eq11}) and (\ref{eq:funnel}). In the simulation, we set $N=50$ (for preparation of Fig. \ref{fig2}). The initial state of the system is usually a node of the graph namely $|\psi_{\rm in}\rangle = |x \rangle$. We represent it by a vector of dimension $N$. For example, if the system is initially localized on node 0, we set the first entry of the vector to be one and all the rest remains zero. With the initial state and funnel/crawl Hamiltonians, we numerically calculate $\phi_1$, which is the overlap between the wave function at time $\tau$ and the search target $|\psi_{\rm d}\rangle$, namely $\phi_1 = \langle \psi_{\rm d}|U(\tau)|\psi_{\rm in}\rangle$. The square of $|\phi_1|$ is the probability that we detect the particle in the first measurement at time $t=\tau$, which is recorded for plotting Fig. \ref{fig2}. We then turn to the calculation of $F_2$. In the first step, the failed measurement (at time $\tau$) projects out the state on $|\psi_{\rm d}\rangle$. This is done by setting the state that overlap with $|\psi_{\rm d}\rangle$ to zero, in other words, we mimic the back-action of projection $(1-|\psi_{\rm d}\rangle \langle \psi_{\rm d}|)$. For example, let $|\psi_{\rm d}\rangle = |0\rangle$, then after the measurement, the state of the system on node 0 is zero. The measured state is the new initial state for the calculation of $F_2$. Similar to the calculation of $F_1$, we let the system evolve for time $\tau$ by $U(\tau)$, then calculate the overlap between the state of the system and search target, which is $\phi_2$. The search probability $F_2 = |\phi_2|^2$.   Such procedure is repeated, we numerically calculate $F_3, F_4, \cdots, F_n$. The results are plotted in Fig. \ref{fig2} for the crawl ({\bf a}) and funnel ({\bf b}) models. In Fig. \ref{fig:comparison with random H}, we utilize the same process for the calculation of $F_n$ and use the random SK Hamiltonian. The mean measurement times are given by $\langle n \rangle = \sum_{n=1}^{M} n F_n$. In the numerical simulation we choose $M = 100000$. In Fig. \ref{fig:noise tau}, the time interval between two measurements is random, depending on the magnitude of the noises. For each $\tau$, we choose $\tau = (2\pi/N)\{1+ a*uniform[-0.5,0.5]\}$ with $a$ being the magnitude of the noise. For the calculation of the search probability, we use the simulated $F_n$ and sum the first 50 measurements, i.e., $P_{\rm det} = \sum_{n=1}^{n=N=50} F_n$. For each magnitude of the noise, $P_{\rm det}$ is averaged over 1000 realizations.     

E.B. thanks Shimon Yankelevich, Moshe Goldstein, and Lev Khaikovich for comments and suggestions. 
The support of Israel Science Foundation's grant 1614/21 is acknowledged.

\clearpage
\newpage
\clearpage 
\setcounter{equation}{0}%
\setcounter{figure}{0}%
\setcounter{table}{0}%
\renewcommand{\thetable}{S\arabic{table}}
\renewcommand{\theequation}{S\arabic{equation}}
\renewcommand{\thefigure}{S\arabic{figure}}
\renewcommand{\thesection}{S\arabic{section}}
\newtheorem{proposition}{Proposition}
\onecolumngrid

\begin{center}
    {\Large Supplemental Information for\\ ``Designing exceptional-point-based graphs yielding topologically guaranteed quantum search''}
    
    \vspace*{0.5cm}
    
    Quancheng Liu, David A. Kessler, and Eli Barkai \\

    \vspace{0.5cm}
\end{center}

\section{Figure preparation}
We now provide further details of the figures.

\begin{itemize}
	\item Fig. 1({\bf a}). Schematic plot of the crawl graph Hamiltonian Eq. (16) of size $20 \times 20$. Here we set $\gamma=1$ and subsequently.  In Eq. (16), we have a typical matrix element $1/[1-\exp(i \theta)]$, which can be formally written as $ 1/[1-\exp(i \theta)] = R \exp(i \Phi)$. The $R$ represents the coupling strength between the nodes, in the figure we utilize the thickness of the connecting line to represent its magnitude. The $R$ decreases as the distance between the nodes becomes longer. For example, $R_{0,1} = R_{0,19} \textgreater R_{0,2} = R_{0,18} \textgreater R_{0,3} = R_{0,17} \textgreater \cdots \textgreater R_{0,10}$. The colors represent the phases $\Phi$, where the phases   $\pi \textgreater \Phi \textgreater \pi/2$. The on site energy of the nodes are equal to zero, hence all of them are plotted gray. Geometrically, the system is rotationally invariant.
	\item Fig. 1({\bf b}). Schematic plot of the funnel Hamiltonian Eq. (17) of size $20 \times 20$. The matrix elements are real, and again we unitize the thickness of the line connecting the nodes to represent the hopping rate magnitude. Now the on-site energies are not identical, and we present this variation by the colors. The detection node is specialized for this model, and we mark this node in the graph. The on-site energies increase linearly for the nodes from 1 to $N-1$, and for the search node, the on-site energy is approximately $N/2$. 
	\item Fig. 3({\bf a}). Search probability $F_n$ versus measurement steps $n$ for the crawl graph. Here the graph has 50 nodes. We choose the initial states to be the nodes of the graph, namely $|\psi_{\rm in}\rangle=|0\rangle,|1\rangle,\cdots, |49\rangle$.  As we discussed in the main text, the search state can be any node of the graph, but here for demonstration, we choose $|\psi_{\rm d}\rangle = |0\rangle$. We numerically simulate the search process with the method we discussed above.  As shown in the figure, the search is deterministic, namely, we detect the walker with probability one at some specified times, as described in the text. 
	\item Fig. 3({\bf b}). Search probability $F_n$ versus measurement steps $n$ for the funnel model. Here we set $N=50$ and the search target is $|\psi_{\rm d}\rangle =|0\rangle$. Again, we choose the initial state to be a node of the graph and $x$ goes from 0 to 49. We then apply the simulation approach discussed above, which gives the statistics of $F_n$ as shown in the figure. For any initial state, the detection of the state is guaranteed with probability one within $N$ measurements. There is a clear cutoff for $F_n$ when $n \textgreater N$, which drops to zero, namely $F_{n \textgreater N}=0$. The upper bound of the search time is found when $|\psi_{\rm in}\rangle =|0\rangle =|\psi_{\rm d}\rangle$, where $F_{50}=1$ and $F_{n \neq 50} =0$, and in this case the detection time is $2\pi$.
	\item Fig. 6 describes the unitary evolution without measurements (non-monitored quantum walks) for the crawl Hamiltonian ({\bf a}) and funnel model ({\bf b}). We plot the probability of finding the walker on node $x$, $|\langle  \psi(t)|x\rangle|^2$ versus $x$, and for continuous-time $t$ (in the unit $\tau/10$). Here, for both graphs, we choose $N=20$ and the initial state is $|0\rangle$. We record $|\langle  \psi(t)|x\rangle|^2$ for all the nodes with sampling time interval $\tau/10$. As shown in the figure, the wave function of the crawl graph is localized at specific nodes of the graph at times $\tau, 2 \tau, 3 \tau, \cdots$. In the funnel mode ({\bf b}), starting for a localized state $|0\rangle$, the wave function first spreads to the whole graph. Then it returns to the localized state $|0\rangle$. The system is recurrent, which is rooted in the periodicity of the energy spectrum we design.
\end{itemize}

\section{DETAILS ON THE DERIVATION OF EQ. 6}

We present the derivation of Eq. (6). As discussed in the main text, the eigen function of the survival operator ${\cal S}(\tau)$ can be written as $\xi \sum_{k=0}^{N-1} p_k /[\xi - \exp(-i E_k \tau) ] = 0$. Here, we denote the summation as $I$. With Eq. (5), we have
\begin{equation}
	I = \frac{\xi}{N} \sum_{k=0}^{N-1} \frac{1}{\xi - \exp(-i 2 \pi k/N) }= - \frac{\xi}{N} \sum_{k=0}^{N-1} \frac{ \exp(i 2 \pi k/N )}{1- \xi \exp(i 2 \pi k/N)},
\end{equation}
where we multiply both the numerator and denominator by $\exp(-i 2 \pi k/N)$ for each term in the summation. We first Taylor expand $1/[1- \xi \exp(i 2 \pi k/N)]$ and get
\begin{equation}
	I = - \frac{\xi}{N} \sum_{k=0}^{N-1} \exp(i 2 \pi k/N ) \sum_{j=0}^{\infty} [ \xi \exp(i 2 \pi k/N) ]^j = - \frac{\xi}{N} \sum_{k=0}^{N-1} \sum_{j=0}^{\infty} \xi^j \exp[i 2 \pi k (j+1)/N].
	\label{smeq05}
\end{equation}
We then calculate $I$ by changing the order of the summations. Namely we first perform the summation over $k$, which is a geometric progression with common ratio $\exp[i 2 \pi (j+1) /N ]$. By calculating the geometric progression, we have:
\begin{equation}
	I= - \frac{\xi}{N} \sum_{j=0}^{\infty} \xi^j \sum_{k=0}^{N-1} \exp[i 2 \pi k (j+1)/N] = \frac{\xi}{N} \sum_{j=0}^{\infty} \xi^j \frac{1- \exp(i 2 \pi j) }{1- \exp[ i 2 \pi (j+1)/N] }.
	\label{smeq06}
\end{equation}
Since $j$ is an integer, the numerator $1-\exp(i 2 \pi j )$ always equals to zero. The whole fraction is non-zero only when the denominator $ 1- \exp[ i 2 \pi (j+1)/N]$ also equals to zero. That is possible and happens when $ \exp[ i 2 \pi (j+1)/N] = 1$, namely $j = n N-1$, where $n$ is an integer and goes from 1 to infinity (if $n$ starts from 0, $j=-1$, which goes beyond the regime of $j$). We replace the summation index $j$ with $n$, where $n$ goes from 1 to infinity. Then for the summation $I$, we have:
\begin{equation}
	I= \frac{\xi}{N} \sum_{j = n N -1   } \xi^j N = \frac{\xi}{N} \sum_{n=1}^{\infty} \xi^{n N -1} N = \sum_{n=1}^{\infty} \xi^{n N } = -\frac{\xi^N}{1-\xi^N}.
\end{equation}
These are the details of the derivation of Eq. (6) in the main text.

\section{Necessary condition for the $N$-th order exceptional point}
In this section, we will show Eq. (5) derived in the main text is a necessary condition for the $N$-th order exceptional point. For following Eq. (3) in the main text, the eigenvalue function for $\xi$ reads:
\begin{equation}
	{\cal F}(\xi) = \langle \psi_{\rm d}| \frac{1}{\xi-U(\tau)}|\psi_{\rm d}\rangle = \sum_{k=0}^{N-1}\frac{p_k}{\xi -e^{-iE_k \tau} }=0. 
\end{equation}
We now prove Eq. (5) in the main text is the only solution to  have a degeneracy of $N-1$ for $\xi_0 =0$. Namely, this equation is a necessary condition for the high order exceptional point we derived. Mathematically, when $\xi_0 =0$ is $N-1$ degenerate, we have
\begin{equation}
	{\cal F}(\xi_0) =0,\  {\cal F}^{\prime} (\xi_0)=0, \ {\cal F}^{\prime \prime} (\xi_0)=0, \  {\cal F}^{(3)} (\xi_0)=0, \ \cdots, \ {\cal F}^{(N-2)} (\xi_0)=0,
	\label{apn1}
\end{equation}
where $\prime$, $\prime \prime$, and $(3)$ denote the first, second, and third order derivative. These conditions leads to $N-1$ equations for the $p_k$, $E_k$, and $\tau$. Since $p_k = | \langle E_k|\psi_{\rm d}\rangle |^2$, we also have the conditions for the form of the $p_k$s, where
\begin{equation}
	\sum p_k =1, \quad \forall k, \ p_k \ \text{is real and positive}. 
	\label{apn2}
\end{equation}
Eqs. (\ref{apn1}) and (\ref{apn2}) ensure Eq. (5) derived in the main text is a necessary condition for the exceptional point. To see that, let us start with the simple case when $N=2$. Using Eqs. (\ref{apn1}) and (\ref{apn2}), the function for the $p_1$ and $p_2$ are
\begin{equation}
	p_1 e^{i E_1 \tau} +(1-p_1) e^{i E_2 \tau} = 0 \ \longrightarrow\ p_1 = \frac{e^{i E_2 \tau} }{e^{i E_2 \tau}-e^{i E_1 \tau}} = \frac{e^{i \Delta_{21} } }{e^{i\Delta_{21} }-1}, \quad p_2 =1- p_1.
	\label{apn3}
\end{equation}
Here we define the energy difference times $\tau$ as $\Delta_{21}$, i.e., $(E_2-E_1)\tau = \Delta_{21}$. Since $p_1$ is real and finite [Eq. (\ref{apn2})], we have $ e^{i\Delta_{21} } =-1$ in the complex plane. This gives the conditions for the energy spectrum, i.e., $ \Delta_{21} = (E_2-E_1)\tau = \pi + 2 k \pi, k \in Z$. Namely the phase between $E_2\tau$ and $E_1\tau$ is $\pi$, as given in Eq. (5). Put the energy spectrum back into Eq. (\ref{apn3}), we have $p_1 =1/2$ and $p_2 = 1/2$, the equal magnitude as we presented in the main text. So for the $N=2$ case, the only solution for the degenerate exceptional point is when $p_1 = p_2 =1/2$ and $(E_2-E_1)\tau = \pi + 2 k \pi$.

Similarly, for $N=3$, we have
\begin{equation} 
	\{ \begin{aligned}
	&p_1 e^{i E_1 \tau} + p_2 e^{i E_2 \tau} + (1-p_1-p_2)e^{i E_3 \tau} =0\\
	&p_1 e^{2i E_1 \tau} + p_2 e^{2i E_2 \tau} + (1-p_1-p_2)e^{2i E_3 \tau} =0
	\end{aligned}, 
	\rightarrow p_1 = \frac{e^{i(\Delta_{21}+\Delta_{31})}}{(-1+e^{i \Delta_{21}}) (-1+e^{i \Delta_{31}})}, p_2 = \frac{e^{i\Delta_{31}}}{(e^{i \Delta_{21}}-1) (e^{i \Delta_{31}}-e^{i \Delta_{21}})}. 
	\label{eqap13}
\end{equation}
Here $\Delta_{21}=(E_2-E_1)\tau$ and $\Delta_{31}=(E_3-E_1)\tau$. Using the conditions in Eq. (\ref{apn2}), we have 
\begin{equation}
	\Delta_{21} = \frac{2\pi}{3} +2 k_1 \pi, \quad \Delta_{31} = \frac{4\pi}{3} +2 k_2 \pi, \quad k_1,k_2 \in Z.
	\label{eqap12}
\end{equation}
This is the energy spectrum condition we have in the main text. For the magnitude of the $p$s, substituting Eq. (\ref{eqap12}) back into Eq. (\ref{eqap13}),  we have $p_1 = p_2 = p_3 =1/3$. 

So in general, for the $N$ dimensional system, using Eqs. (\ref{apn1}) and (\ref{apn2}), we have
\begin{equation}
	p_1 = \frac{e^{i \sum_{i=2}^{N} \Delta_{i1} }}{\Pi_{i=2}^{N} (e^{i\Delta_{i1}-1} )}, p_2 = \frac{e^{i \sum_{i=3}^{N} \Delta_{i1} }}{(e^{i\Delta_{21}}-1) \Pi_{i=3}^{N} (e^{i\Delta_{i1}- e^{i\Delta_{31}}} )}, \cdots, p_k = \frac{e^{i \sum_{i=2, i\neq k}^{N} \Delta_{i1} }}{(e^{i\Delta_{k1}}-1) \Pi_{i=2, i\neq k}^{N} (e^{i\Delta_{i1}- e^{i\Delta_{k1}}} )}.
	\label{apnfvg1} 
\end{equation}
Since these $p_k$ should be real and possible, we have the conditions for the $\Delta_{i1}$s. This process leads to the energy spectrum conditions as presented in Eq. (5). Substituting the energy level conditions back into Eq. (\ref{apnfvg1}). The corresponding magnitude of the $p_k$ are all equal, i.e., $p_k = 1/N$. To conclude, Eq. (5) is a necessary condition for achieving an $N$-th order exceptional point.

\section{Proof of the orthogonal of states $|Q_k \rangle$}
We present the proof that the states $|Q_0 \rangle, |Q_1 \rangle, |Q_2 \rangle, \cdots, |Q_{N-1} \rangle$ are orthogonal with each other, namely $\langle Q_l | Q_m \rangle =\delta_{lm}$. The state $|Q_k \rangle$ is defined by the unitary evolution operator $U_s$ to the power $k$ times the search target $|\psi_{\rm d}\rangle$, where $U_s = \exp(-i H_s \tau)$ and $H_s$ are search Hamiltonian. To show the orthogonality, we first expanded $|\psi_{\rm d}\rangle$ in the energy basis, which leads to:
\begin{equation}
	|Q_m \rangle = (U_s)^m |\psi_{\rm d}\rangle = \sum_{k=0}^{N-1} (U_s)^m \langle E_k | \psi_{\rm d}\rangle |E_k\rangle = \sum_{k=0}^{N-1} \exp(-i m E_k \tau) \langle E_k | \psi_{\rm d}\rangle |E_k\rangle = \sum_{k=0}^{N-1} \exp(-i 2 \pi k m/N) \langle E_k | \psi_{\rm d}\rangle |E_k\rangle.
\end{equation}
Here we have used the fact  $E_k \tau = 2 \pi k/N$. Similarly, we can find the representation of the state $ |Q_l \rangle $ in the energy basis. We then calculate the overlap between the states $|Q_m\rangle$ and $|Q_l\rangle$. We have
\begin{equation}
	\langle Q_l | Q_m \rangle = \sum_{k=0}^{N-1}  \sum_{k^{\prime}=0}^{N-1} \langle E_k| \langle\psi_{\rm d}|E_k\rangle \exp[i 2\pi (k l-k^{\prime} m)/N ]\langle E_{k^{\prime}}|\psi_{\rm d}\rangle |E_{k^{\prime}}\rangle = \sum_{k=0}^{N-1} |\langle\psi_{\rm d}|E_k\rangle|^2 \exp[i 2\pi k (l - m)/N].
	\label{smeq10}
\end{equation}
The square of the overlap between the detected state $|\psi_{\rm d}\rangle$ and the energy state $|E_k\rangle$ is denoted as $p_k$ in the main text, namely $p_k =|\langle\psi_{\rm d}|E_k\rangle|^2$. Eq. (5) states this value is a dependent of $k$ for the search Hamiltonian $H_s$, where $p_k = 1/N$. So for Eq. (\ref{smeq10}), we only need to calculate the summation of $\exp[i 2\pi k (l - m)/N]$ from $k=0$ to $N-1$. This has been done also in Eq. (\ref{smeq06}), where $(j+1)$ in Eq. (\ref{smeq06}) is replaced by $(l-m)$ here. We then have:
\begin{equation}
	\langle Q_l | Q_m \rangle = \frac{1}{N} \sum_{k=0}^{N-1} \exp[i 2\pi k (l - m)/N] = \frac{1}{N} \frac{\exp[i 2 \pi (l-m)]-1}{ \exp[i 2 \pi (l-m)/N]-1 }.
\end{equation}
$\langle Q_l | Q_m \rangle$ is non-zero only when $l-m = 0,N,2N, \cdots$. Here $N-1 \geq l \geq 0$ and $N-1 \geq m \geq 0$. Hence only when $l=m$ we have $\langle Q_l | Q_m \rangle =1$, otherwise $\langle Q_l | Q_m \rangle =0$. Namely $\langle Q_l | Q_m \rangle = \delta_{lm}$.
%
%
This is the conclusion we used in the main text. Another thing to notice is that, since the $|Q_k\rangle$ are generated by the unitary operators, it is naturally normalized. Hence the states $\{|Q_0\rangle, |Q_1\rangle, |Q_2\rangle, \cdots, |Q_{N-1}\rangle \}$ forms a complete and normalized space.

\section{Funnel Model Hamiltonian}
We provide details on the funnel Hamiltonian and its explicate presentation. In this model, the search target $|\psi_{\rm d}\rangle =|0\rangle$. As before, the spatial nodes of the graph are denoted $|x_i\rangle$, and $i=0,1, \cdots, N-1$. We start with the  energy state $|E_0\rangle = \left\{1/\sqrt{N},-\sqrt{(N-1)/N}, 0, 0, \cdots \right \}$, where $1/\sqrt{N}$ fulfills the first condition in Eq. (5) and $-\sqrt{(N-1)/N}$ stands for normalization. We then construct the energy state $|E_1\rangle$. It should be orthogonal with $|E_0\rangle$ and in agreement with the condition in Eq. (5). We find $|E_1\rangle = \left\{1/\sqrt{N}, 1/\sqrt{N(N-1)}, -\sqrt{(N-2)/(N-1)}, 0, 0, \cdots \right \}$. Again, the first term in $|E_1\rangle$ leads to $|\langle E_1|0\rangle|^2=1/N$. The second entry guarantees $\langle E_0|E_1\rangle =0$, and the third entry is for normalization. Following the construction procedure, we have $|E_2\rangle = \left\{1/\sqrt{N}, 1/\sqrt{N(N-1)}, 1/\sqrt{(N-2)(N-1)}, -\sqrt{(N-3)/(N-2)}, 0, 0, \cdots \right \}$, and in general $|E_{i \neq N-1} \rangle = \left\{1/\sqrt{N}, 1/\sqrt{N(N-1)}, \cdots, 1/\sqrt{(N+2-k)(N+1-k)}, \cdots,-\sqrt{(N-1-i)/(N-i)}, 0, 0, \cdots \right \}$, where $k$ is the index for the $k$-th entry and $ 2\leq k \leq i$. This general representation $|E_i\rangle$ can help us construct the states $|E_0\rangle, |E_1\rangle, \cdots, |E_{N-2}\rangle$, but not $|E_{N-1}\rangle$. Let us explain this and then show how to construct $|E_{N-1}\rangle$. When $i=N-2$, we have $|E_{N-2}\rangle = \left\{1/\sqrt{N}, 1/\sqrt{N(N-1)}, \cdots, 1/\sqrt{6}, -1/\sqrt{2}  \right \}$. Now we cannot utilize the procedure we did before to construct $|E_{N-1}\rangle$, since roughly speaking there is no additional space for the normalization. So how to construct the last energy state? We notice the last term in $|E_{N-2}\rangle$ is $-1/\sqrt{2}$, which is special. Let us consider a state where the first $N-1$ terms are the same as those in $|E_{N-2}\rangle$, and the only different one is the last term, where we change the $-1/\sqrt{2}$ to $1/\sqrt{2}$. We can see such a state is orthogonal with $|E_{N-2}\rangle$ and also normalized. This is the last energy state, i.e.,  $|E_{N-1}\rangle = \left\{1/\sqrt{N}, 1/\sqrt{N(N-1)}, \cdots, 1/\sqrt{6}, 1/\sqrt{2}  \right \}$. It is also easy to show that this state is orthogonal with respect to the other state, $|E_{N-3}\rangle, |E_{N-4}\rangle, \cdots, |E_0\rangle$.  We then have $N$ orthogonal and normalized states fulfill the conditions in Eq. (5).

For the equal distance energies, we set $E_0 =0, E_1= \gamma, E_2 = 2\gamma, \cdots, E_{N-1}=(N-1)\gamma$, the resulting Hamiltonian is 

\begin{equation}
H = {\gamma \over 2}  \left[
\begin{matrix}
N-1 & H_{0,1} & H_{0,2} &  \cdots &  H_{0,N-1}  \\
H_{0,1}  & 1 & H_{1,2} &  \cdots  & H_{1,N-1}  \\
H_{0,2}  & H_{1,2}  & 3 &  \cdots  & H_{2,N-1}  \\
 \vdots & \vdots  & \vdots   & \ddots  & \vdots \\ 
H_{0,N-1} & H_{1,N-1} & H_{2,N-1} &    \cdots & 2N -3 
\end{matrix}
\right]
\label{apeq12}
\end{equation}
where $H_{0,m}=\sqrt{(N-m)(N-m+1)/N}$ ($m\neq 0$) and $H_{j, m}=\sqrt{(N-m)(N-m+1)/[(N-j+1)(N-j)]}$ ($j \neq 0,m$). We call this approach the funnel model.

\end{document}